\documentclass[%
reprint,
amsmath,amssymb,
aps,
pra,
]{revtex4-2}

\usepackage{graphicx}% Include figure files
\usepackage{dcolumn}% Align table columns on decimal point
\usepackage{bm}% bold math
\usepackage{tikz}
\usepackage{pgfplots}
\usepackage{float}
\newcommand{\ket}[1]{\left|{#1}\right\rangle}

\newcommand*{\avg}[1]{\langle{#1}\rangle}

\pgfplotsset{compat=1.17}

\begin{document}

%\preprint{APS/123-QED}

\title{Topological two-particle dynamics in a periodically driven lattice model with on-site interactions }

\author{Anna Berti}
%\email{anna.berti-1@unitn.it}
\author{Iacopo Carusotto}
%\email{iacopo.carusotto@unitn.it}
\affiliation{ INO-CNR BEC Center and Dipartimento di Fisica, Universit\`a di Trento, 38123 Trento, Italy }

\date{\today}

\begin{abstract}
	We develop a realistic protocol to observe a robust topological dynamics of two-particle bound states in a lattice model with on-site interactions and suitably designed time-dependent hoppings. This Floquet scheme can be realistically implemented on existing digital quantum computer platforms. Marked differences from the topological single-particle dynamics of two independent particles and clear signatures of the entanglement between the two constituent particles are highlighted.
\end{abstract}

\maketitle

\section{Introduction}
Since their discovery, quantum Hall effects in two-dimensional electron gases under strong magnetic fields have been the subject of many theoretical and experimental studies highlighting their strong connection with topology concepts~\cite{iqhe1,fhqe1,tong,tknn,fqhe2,fqhe3, Bernevig2013, franz2013topological}: in particular the quantization of the Hall conductance is strictly related to the integer nature of a topological invariant known as Chern number, characterizing the geometry of the Bloch state wavefunctions across the Brillouin zone.

Besides electrons in solid-state systems, the topological single-particle dynamics underlying the integer quantum Hall effect has been experimentally investigated in several other systems, including cold atomic gases~\cite{review_dalibard_cooper_spielman} and photonic platforms~\cite{topphot}. 
For example, the quantized transverse motion has been observed using atoms in an optical lattice in~\cite{aidelsburger2015}, while chiral edge states have been seen in ladder configurations~\cite{atala2014} as well as in a synthetic dimension platform~\cite{mancini2015,stuhl2015}. Concerning photonic systems, unidirectionally propagating chiral edge states have been seen in a variety of material systems starting with~\cite{haldane, wang2009} and anomalous Hall effects were observed in~\cite{wimmer2017}.
In the presence of interactions, a much richer physics is found in the strongly correlated states underlying the fractional quantum Hall effect, including fractional charge and statistics of anyonic excitations~\cite{stern} and exciting perspectives in the direction of topological quantum computing~\cite{topoqc}.
Beyond two-dimensional electronic gases in different materials under strong magnetic fields~\cite{tong}, first steps towards the observation of these phenomena in cold atomic gases in optical lattices~\cite{Preiss_2015,Tai_2017}, arrays of Rydberg atoms trapped in optical tweezers~\cite{deLese_2019}, Rydberg polaritons in atomic gases~\cite{clark2020observation}, and superconductor-based circuit-QED systems~\cite{Roushan2016,Ye_2019,RevModPhys.93.025005} have been recently reported. 

With the recent progress on digital quantum computing (QC) devices, platforms hosting a lattice of qubits and allowing for discrete single- and two-qubit gate operations between neighboring sites~\cite{google,nielsen2010quantum, surveyQC} appear as promising platforms to explore many-body phenomena. In these QC platforms, discrete unitary operations are applied in a sequence, so to guarantee that each site is subject to a single operation during each time-step. 
Although this is a necessary step in the optimization of a QC device, it is also responsible for the typical complication of dealing with time-dependent Hamiltonians, typically chosen with a periodic Floquet form. 

Of course, static models can be simulated on these Floquet platforms by performing a procedure called \textit{trotterization}: the Hamiltonian $\mathcal H$ is split as the sum of several (typically non-commuting) contributions $\mathcal H_n$ and the time-evolution operator $\exp(-i\mathcal H \tau)$ is computed as the product of the exponential $\exp(-i\mathcal H_n \tau)$ of each operator. According to the Trotter product formula~\cite{nielsen2010quantum}, such approximation is accurate also for non-commuting $\mathcal{H}_n$ if the time-step $\tau$ is small with respect to all the energy scales of the problem, so that the terms involving the commutators, of higher order in $\tau$, are negligible. In physical terms, any static model can be realized through a Floquet evolution if operations are performed at a fast enough rate, so that their ordering becomes irrelevant. If this procedure is done in an accurate way by taking a small enough $\tau$, we speak of \textit{good trotterization}.  

However, since in any realistic device only a limited number of gates can be performed before losing coherence it is often beneficial to adopt sizable unitary operations well outside the good trotterization regime, a regime in which the Floquet nature of the dynamics can have a substantial impact ~\cite{floquet,goldman2014periodically,rudner}. 
As an illustration of the potential and the complexities of this approach, this work reports a detailed investigation of the topological Floquet dynamics of two-particle bound states, the so-called doublons~\cite{Winkler_2006,petrosyan2007quantum,Valiente_2008,folling2007direct}. In spite of energy being conserved only modulo the Floquet energy, suitable working points are found where doublon states can not dissociate into free particles and a clean unidirectional propagation of doublons around the system edge is anticipated. In its simplicity, such a two-body effect is an example of the subtle interplay between topology and interactions \cite{Qin2017, Ke2017, Qin_2018,Gorlach_2017,doublons1,doublons2,molecules,Azcona_2021,Ke2020, Lin2020} and a preliminary step towards the realization of topological many-body states of photonic matter~\cite{carusotto2020photonic}.

The structure of the article is the following. Sec.\ref{sec:model} introduces the physical system and the theoretical model. The topological properties of the Floquet bands in a few most relevant configurations are discussed in Sec.\ref{single}. Our results on the internal structure and the topological propagation properties of two-particle bound states are presented in Sec.\ref{twoparticles}. Further developments in the direction of many-body physics are sketched in Sec.\ref{manybody}. Conclusions are finally drawn in Sec.\ref{conclusion}.

\section{The model}\label{sec:model}
Throughout the work, we consider a two-dimensional bosonic lattice system of time-dependent Hamiltonian
\begin{equation}
\mathcal H(t) = - \sum_{ij} \Big( J_{ij}(t) a_i^\dagger a_j + \text{h.c.} \Big) + \frac U 2 \sum_{i} (a_i^{\dagger})^2 a_i^2
\label{ham}
\end{equation}
where $a_i^\dagger$ ($a_i$) is the creation (annihilation) operator for a particle in the $i$-th site. Interactions are assumed to be on-site with strength $U$ and are active at all time, as implemented in superconducting QC platforms~\cite{Ye_2019}.

With an eye to digital QC frameworks~\cite{google}, the hopping coefficients $J_{ij}(t)$ are instead taken time-periodic of period $T$. 
We assume that each period is split in a sequence of $N$ sub-periods of equal duration $\tau = T/N$. During each sub-period $[(n-1)\tau, n\tau]$ (with $n=1,...,N$) the time-independent hamiltonian $\mathcal{H}(t)=\mathcal{H}_n$ features a constant hopping strength $J$ between a given set of pairs of sites $\avg{ij}_n$ with non-trivial hopping phases, $J_{ij}(t)=J_{ij}^{n}=J\exp(i\phi_{ij}^n)$. 

With these assumptions, the Hamiltonian \eqref{ham} is obtained through a periodic sequence of gate operations, each of them involving only one or two qubits. 
All these operations appear to be well within reach of existing QC platforms: time-dependent hopping gates are realized through adjustable couplings \cite{Chen_2014, Kounalakis_2018}, while interactions and non-trivial hopping phases can be implemented respectively by exploiting non-linear effects \cite{strong_photons2017} and engineering artificial gauge fields \cite{gaugefields2018, Dalibard_2011, Aidelsburger_2011, Goldman_2014}. 

\begin{figure}[htbp]
	\centering
    \includegraphics[width=\linewidth]{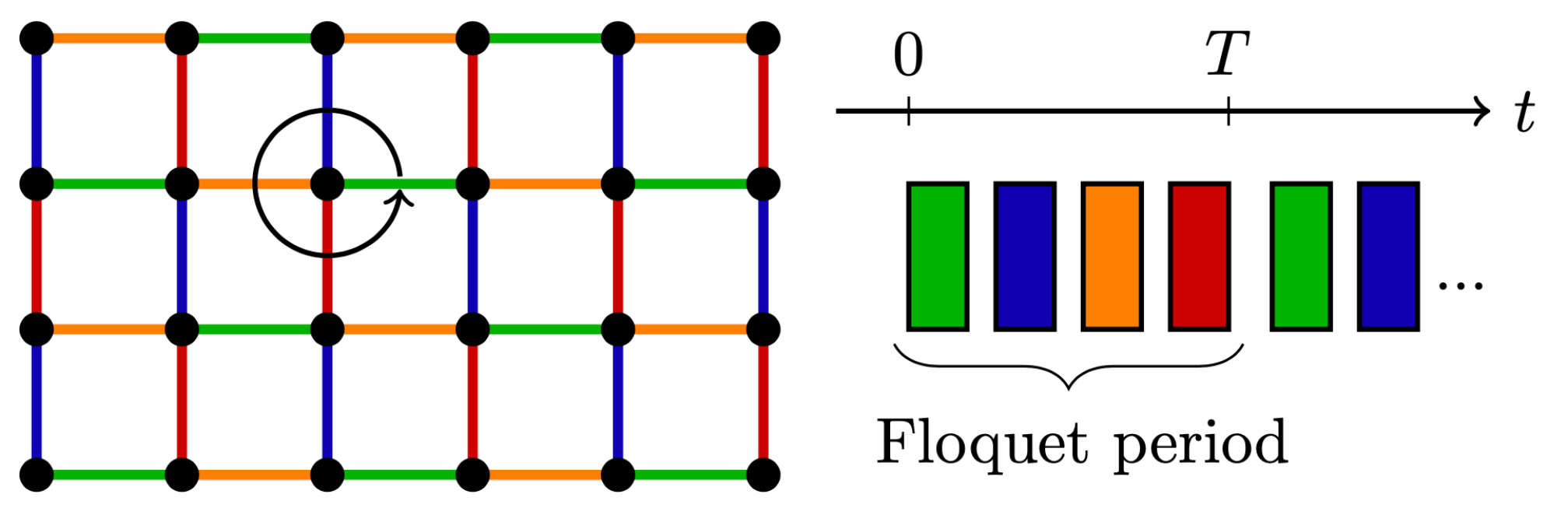}
	\caption{Scheme of the two-dimensional square lattice under investigation. 	Each color identifies the hopping links that are active during each time-interval according to the temporal sequence shown on the right. The arrow indicates the chirality of the hopping scheme.}
	\label{hop}
\end{figure}

The configuration we work with throughout the paper is illustrated in Fig.\ref{hop}. During each sub-interval, only the links of the corresponding color are active (left panel). The order of activation is indicated in the right panel.

The choice of this specific hopping scheme is mainly based on its experimental feasibility: existing devices \cite{google} have square geometry and nearest-neighbors connectivity. An additional advantage of this sequence resides in its intrinsic chirality: for each qubit the links are activated in a counter-clockwise order (of course, if the ordering of steps was reversed, the activation would be in the opposite, clockwise order). As a result, the sequence in Fig.\ref{hop} breaks time-reversal invariance and, as we are going to see in Sec.\ref{single}, is able to transform a model that is topologically trivial in the static case into a topological Floquet one. 

While we focus our attention on this most convenient configuration, analogous results hold for lattices with different geometry (triangular, hexagonal, etc.), connectivity (next-nearest-neighbors, etc.) and ordering of the steps, provided that the hopping scheme only involves single- and two-qubit gates and time-reversal symmetry is effectively broken by the hopping phases and/or the hopping sequence, so that the resulting single-particle dynamics is topological in some range of parameters. 

\section{Topological single-particle physics } \label{single}
We start our investigation by looking at the topological properties of the single-particle Floquet bands arising from our time-dependent Hamiltonian \eqref{ham}. 
Given its discrete temporal periodicity,  within the Floquet theory~\cite{floquet,goldman2014periodically,rudner} the single particle bands arise as the eigenvalues of the Floquet operator, namely the time-evolution operator along a period $T$. 

Since during each sub-interval the Hamiltonian is time-independent, the Floquet operator $\mathcal U(T)$ can be written as the product of the $N$ unitary operators associated to each hopping step, $\mathcal U(T) = \prod_{n=N}^1 \exp(-i \mathcal{H}_n \tau) $. 
Since the pairs $\avg{ij}_n$ of sites involved in each step are disconnected, the exponential $\exp(-i \mathcal{H}_n\tau)$ can be expanded as a block matrix in the single particle basis. The block corresponding to each pair $ij$ is a hopping matrix with hopping angle $\theta = J\tau$,
\begin{equation}
    U_H(\theta,\phi) = e^{iJ\tau (\hat u \cdot \vec\sigma)} = \begin{pmatrix} \cos\theta & i\sin\theta e^{i\phi}\\
    i\sin\theta e^{-i\phi} & \cos\theta \end{pmatrix}
    \label{UH1}
\end{equation}
while the sites not involved in any pair are left intact. 

\begin{figure*}[htbp]
	\includegraphics[height = 0.23\linewidth]{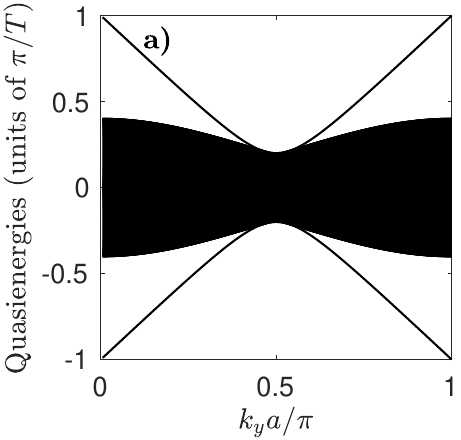}\quad
	\includegraphics[height = 0.23\linewidth]{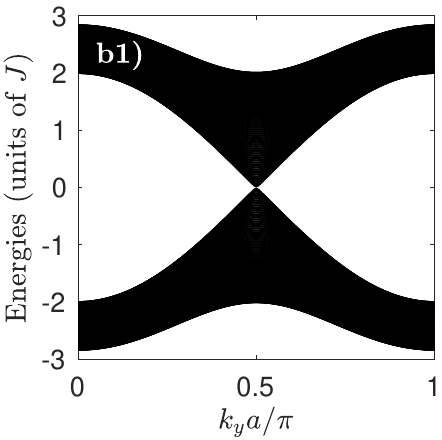}
	\includegraphics[height = 0.23\linewidth]{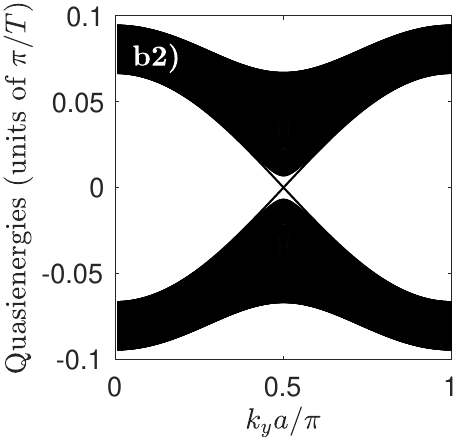}
	\includegraphics[height = 0.23\linewidth]{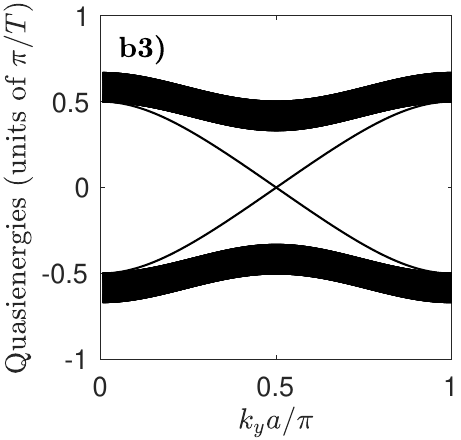}
	\caption{Single-particle spectra computed for a cylidrical geometry with $150$ sites along $x$. From left to right: anomalous Floquet insulator (AFI) model with $\theta = 0.6\pi$; static Harper-Hofstadter model with $\alpha=1/2$; Harper-Hofstadter-Floquet (HHF) model with $\alpha = 1/2$ and $\theta = \pi/30$; HHF model with $\alpha = 1/2$ and $\theta=\pi/4$.}  
	\label{spectra}
\end{figure*}

The good trotterization limit is realized instead when the step size $\tau$ is much smaller than the typical time-scales of the problem, that is $\theta \ll 1$. In this regime, the full Floquet operator $\mathcal{U}(T)$ can be accurately approximated by the exponential of a static Hamiltonian; hence the condition $\theta \ll 1$  allows the realization of most standard time-independent topological models. %, in particular the Hofstadter one.
In this work we go beyond this regime and we look for features that stem from the Floquet character of our dynamics. As anticipated, the topology of the single-particle bands of this Floquet tight-binding model is not only determined by the values of $\theta$ and $\phi_{ij}^n$, but may also depend on the hopping sequence. 
Two most remarkable examples of this physics are illustrated in Fig.\ref{spectra}. 

In Fig.\ref{spectra}(a), we consider the anomalous Floquet insulator (AFI) model of~\cite{rudner}, which features a square lattice with time-dependent nearest-neighbors hoppings and vanishing hopping phases.
This is naturally realized using the hopping sequence described in Fig.\ref{hop} with trivial hopping phases $\phi_{ij}^n = 0$ for all $i,j,n$. For a finite system in the form of a strip with periodic boundary conditions along $y$ and a finite size along $x$, the spectrum involves a single bulk band plus a pair of dispersive edge states on each side that span across the whole quasienergy Brillouin zone.

In Fig.\ref{spectra}(b1-b3) we consider the Harper-Hofstadter model. The time-independent version~\cite{Hofstadter_1976} consists in a square lattice with nearest-neighbors hoppings and non-zero Peierls phase, corresponding to a uniform magnetic field with flux per plaquette $\alpha$. 
This can be realized through a synthetic vector gauge potential along $y$, by imposing $\phi_{ij}^n = \pi x_i$ for all links directed along the positive-$y$ direction. For generic $\alpha$, this model is known to lead to topological bands. If $\alpha = 1/2$ the system is instead topologically trivial: the spectrum shows two bulk bands touching at a Dirac point, as shown in Fig.\ref{spectra}(b1).

In Fig.\ref{spectra}(b2-b3) we show the spectrum of a Harper-Hofstadter-Floquet (HHF) model where the couplings of the standard Harper-Hofstadter model are periodically applied in time according to the sequence shown in Fig.\ref{hop}. In particular, we focus our attention on the most intriguing $\alpha=1/2$ case.
In contrast to the non-topological nature of the underlying static model, the chirality of the chosen hopping sequence leads to the opening of a finite topological gap. This is witnessed by the appearance of topological edge states within the gap and by the onset of finite (and opposite) values of the Chern numbers of the two bands. Of course, the width of the topological gap is only sizable for sizable values of the hopping angle $\theta$ [in Fig.\ref{spectra}(b3) we use $\theta=\pi/4$], and closes down as one approaches the good trotterization regime $\theta\to 0$ recovering the time-independent limit [see Fig.\ref{spectra}(b2)]. 

These properties of the single-particle bands are the starting point of our analysis of the two-body physics.

\section{Topological dynamics of two-particle bound states}
\label{twoparticles}
In the time-independent case, the energy of a doublon is typically of the order of the interaction strength $U$ whereas the continuum of scattering states is characterized by a bandwidth of the order of the hopping coefficients $J_{ij}$. In the strongly interacting limit $|U| \gg |J_{ij}|$, doublons are thus isolated states in the two-particle spectrum and processes in which a doublon decays into its constituents are forbidden by energy conservation~\cite{Winkler_2006,petrosyan2007quantum,Valiente_2008,folling2007direct}: the energy that would be released in such a decay processes is too high to be absorbed by the kinetic energy of the products. As a result, isolated doublons are stable objects and ensembles of many doublons are metastable states that, thanks to kinematic constraints, can only decay or form other many-body objects with exponentially small probability in $|U|/J$ \cite{Mark_2020}. 

In this strongly interacting regime, perturbation theory in $J_{ij}/|U|$ can be used to understand the topological properties of the energy bands of an isolated doublon by deriving an effective single particle Hamiltonian for the two-particle composite object in the usual form,
\begin{equation}
	\mathcal H_\text{eff} = U \sum_{i} \alpha_i^\dagger \alpha_i - \sum_{ij} \Big( J_{ij}^\text{eff}\alpha_i^\dagger \alpha_j + \text{h.c.} \Big)\,
\end{equation}
where $\alpha_i = {a_i a_i}/{\sqrt 2}$ annihilates a doublon on the $i$ site. The effective hopping coefficients $J_{ij}^\text{eff}$ for doublons can be extracted from the ones of the underlying single-particle model by identifying sequences of single-particle hopping processes leading to an overall doublon hopping via virtual intermediate states with spatially separated particles. Such an approach has been used in studies of the band topology of doublons in different lattice geometries for time-independent Hamiltonians~\cite{Qin2017,Ke2017,Qin_2018, Gorlach_2017,doublons1,doublons2,molecules,Azcona_2021,Ke2020, Lin2020}.

\subsection{Floquet two-particle bound states} 

In our periodically driven case, we can still define a doublon as a localized state of two particles but the strong interaction assumption is no longer sufficient to guarantee the stability of doublon states. Since the spectrum of Floquet systems is periodic by $\Omega=2\pi/T$ along the energy direction, one must in fact ensure that the band of doublon states never overlaps with any Floquet replica of the single-particle states. A sufficient condition is of course to work in the trotterized limit of a very fast Floquet frequency $\Omega\gg |U| \gg |J_{ij}|$ where the the Floquet problem can be accurately reformulated in terms of an effective time-independent Hamiltonian.

From a practical point of view, this regime is however not the most favourable one in view of implementations on a digital quantum computer device, as it requires a large number $\sim \Omega/|J_{ij}|$ of elementary operations to be performed during the characteristic time of the particle motion, with the consequent loss of coherence and growth of errors. It is therefore of great interest to devise configurations for which the doublons are stable in regimes where the three energy scales $\Omega,\,U,\,J_{ij}$ are comparable.

Our proposed procedure starts from the computation of the time-evolution operator during a single step of the Floquet evolution in the basis of two-particle states. Consider a generic site $i$ and the time-interval $n$ in which it is coupled to the neighboring site $j$. The three two-particle states coupled by the Hamiltonian $\mathcal H_n$ are indicated as $\ket{ii}$, $\ket{ij}$ and $\ket{jj}$: the first and the third states are doublons localized at the $i,j$ sites respectively; the second state contains a particle in each site. In this basis, exponentiation of $\mathcal H_n$ gives the time evolution operator 
\begin{equation}
	\mathcal U = e^{-i\gamma/2} \begin{pmatrix}
	U_{11} & U_{12}e^{i\phi} & U_{13}e^{2i\phi}\\
	U_{12}e^{-i\phi} & U_{22} & U_{12}e^{i\phi}\\
	U_{13}e^{-2i\phi} & U_{12}e^{-i\phi} & U_{11}
	\end{pmatrix}
	\label{evol1}
\end{equation}
whose matrix elements involve the coefficients
\begin{align}
U_{12} &= i \frac{2\sqrt 2 \theta}{\gamma'} \sin(\gamma'/2) \label{eq:u12}\\
U_{22} &=  \cos(\gamma'/2) + i \frac{\gamma}{\gamma'} \sin(\gamma'/2) \\
U_{11} &= \frac{1}{2} \Big[ U_{22}^* + \exp(-i\gamma/2)  \Big]\\
U_{13} &= \frac{1}{2} \Big[ U_{22}^* -\exp(-i\gamma/2)  \Big] \label{u13}
\end{align}
defined in terms of $\gamma = U\tau$ and $\gamma' = \sqrt{\gamma^2 + 16\theta^2}$.

In order to obtain a closed dynamics within the space of doublon states, we need to initialize the system in a doublon state and then prevent doublons from dissociating into single-occupancy states during the whole evolution. In terms of the time-evolution matrix \eqref{evol1}, this requires that $U_{12}=0$, which in turn imposes:
\begin{equation}
\left(\frac{U}{J}\right)^2 = {\frac{4k^2}{(\theta/\pi)^2} - 16}  \label{uj}  
\end{equation}
for some integer $k$ for which the right-hand-side is positive. As usual for doublons, the properties of the system are invariant under $U\rightarrow -U$. Hence, we can assume $U > 0$ with no loss of generality. 

Under the condition \eqref{uj}, the second row and column of the time evolution matrix decouples from the two other and we can recast \eqref{evol1} in terms of a two-by-two hopping matrix for doublon states in the $i,j$ sites analogous to \eqref{UH1}
with effective hopping angle $\theta'$ and phase shift $\phi'$
\begin{eqnarray}
\theta' &=&  \frac \pi 2 \left( \text{mod}(k,2) + \sqrt{k^2-(2\theta/\pi)^2}\right) \label{thetaprimoo}\\
\phi' &=& 2\phi\,. \label{phiprimoo}
\end{eqnarray}

Combination of Eqs.(\ref{uj}-\ref{phiprimoo}) then fully characterize the two-particle model: the first provides the specific values of $U/J$ for which the two-particle dynamics can be reduced to an effective one-body dynamics for doublons. The second and the third give the values of the parameters $\theta',\phi'$ of the effective model, from which it is straightforward to establish whether the energy bands of doublons have non-trivial topological properties.
As expected, in the $\theta\ll 1$ good trotterization limit  corresponding to an effectively static model, Eq.\eqref{uj} is only satisfied with strong interactions $|U| \gg J$.

\begin{figure}[htbp]
	\centering
	\includegraphics[height=0.63\linewidth]{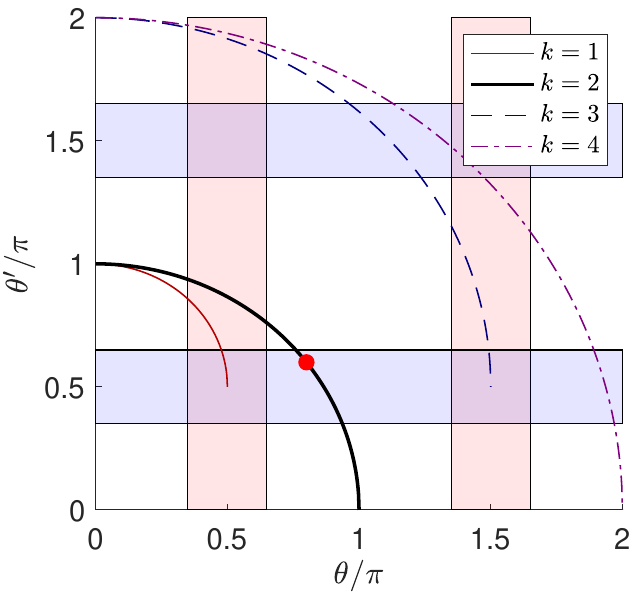}
	\includegraphics[height=0.62\linewidth]{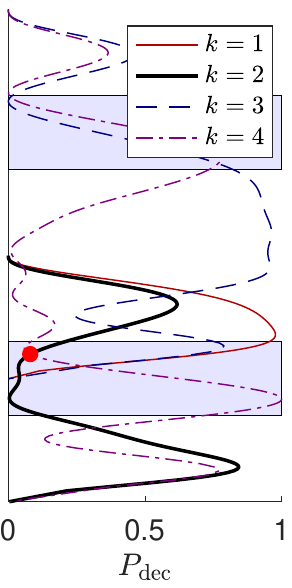}
	\caption{Left panel: functions $\theta'(\theta)$ given by \eqref{thetaprimoo} and defining the stability of doublon states, for different values of $k$ and $U>0$. Red (blue) areas indicate a non-trivial band topology for single-particle (doublons) for trivial hopping phases $\phi_{ij}=\phi'_{ij}=0$; they are estimated by looking for the presence of edge states. 
	Right panel: probability that two doublons occupying coupled neighbouring sites remain intact at the end of a hopping step as a function of $\theta'$ for different $k$ and assuming $U> 0$.
	The red dots indicate the parameter values adopted in the following Figures.}
	\label{thetaprimo}
\end{figure}

\begin{figure*}[htbp]
	\centering
	\includegraphics[width=\linewidth]{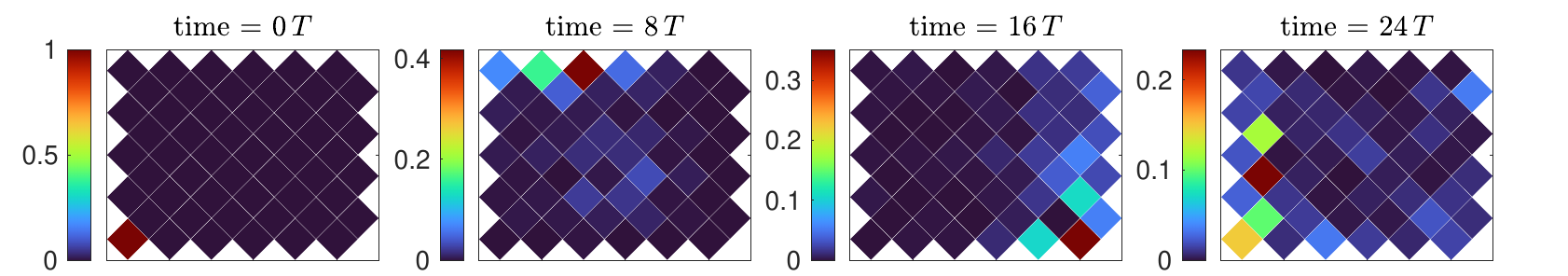}
	\caption{Numerical simulation of the time-evolution on a $N_\text{sites}=54$ sites squared lattice. For each site, the color indicates the probability of having a doublon at that position, $A_l=|\avg{\psi|ll}|^2$. System parameters: $\phi_{ij}=0$, $\theta = 0.8\pi$, $U=3J$ as indicated by the red dot in Fig.\ref{thetaprimo}. The system is initialized with a doublon on the bottom-left corner. The doublon then evolves in the clock-wise direction around the system.} 
	\label{topoevol}
	\end{figure*}

\subsection{Topological two-particle dynamics}

To illustrate the impact of the Floquet dynamics, we focus on a most relevant specific example based on an anomalous Floquet insulator configuration with trivial hopping phases $\phi=0$ for all pairs of sites. The curves in Fig.\ref{thetaprimo} show the effective two-particle hopping angle $\theta'$ given by \eqref{thetaprimoo} as a function of the single-particle hopping angle $\theta$ for the stable two-particle bound states fulfilling \eqref{uj} with  $k=1,2,3,4$. According to \eqref{phiprimoo}, the two-particle hopping phase is also trivial $\phi'=0$. The red and blue shadings respectively indicate the parameter ranges for which the one-body and two-body dynamics are topological. 
Quite interestingly, this plot shows the existence of ranges of parameters in which the doublon band may display non-trivial topological properties while the underlying single-particle model is topologically trivial (and viceversa).
Let us now investigate the most promising choices of microscopic parameters $U,J,T$ to achieve these regimes.

In the strongly interacting limit $|U|\gg J$, the decoupling condition \eqref{uj} can be satisfied with $k\gg 1$ and the effective hopping angle for doublons  is (modulo $\pi$) $\theta' \sim -2J\theta/|U| =  (-2J^2/|U|)\tau$. This corresponds to an effective hopping coefficient $J_\text{eff}=-2J^2/|U|$ equal to the one that one would get for a static system using perturbation theory. This conclusion holds under no assumption on the value of $\theta$ and proves that, independently of the presence or absence of the periodic driving, the dynamics of doublons is always very slow in the strongly interacting limit, $|\theta'| \ll \theta$. Once again, this is not a convenient regime for an experiment, since it requires a large number of gate operation at the cost of introducing significant errors. 
As a consequence, all the calculations that follow will focus on the smallest values of $U$ that are solutions of the decoupling condition \eqref{uj}. 

In particular, we focus our attention on the configuration with  $\theta = 0.8\pi$ and $U = 3J$ (and $\phi=\phi'=0$). This parameter choice (indicated by the red dot in Fig.\ref{thetaprimo}) satisfies \eqref{uj} with $k=2$ and gives a sizable value of the effective hopping angle $\theta' = 0.6\pi$ for which the two-body evolution is topological [see Fig.\ref{spectra}(a)]. 

The time-evolution starting from a doublon state localized in the bottom-left corner is reported in Fig.\ref{topoevol}: as expected, the spatial distribution of the double-occupancies $A_l=|\avg{\psi|ll}|^2$ unidirectionally circulates around the edge while spreading in space because of the non-zero group velocity dispersion of the doublon edge state. As typical for a chiral dynamics, the wavepacket is able to turn around corners without suffering any backscattering. Given the non-topological nature of the single-particle bands for $\theta = 0.8\pi$, no propagation on the edge would be observed in the absence of interactions $U=0$: the two particles would simply diffuse in space independently penetrating the bulk. Along the same line, the evolution would also be trivial in the static limit $\theta \ll 1$, because the topology of the AFI model is entirely due to the periodic driving.

It is worth noticing that, for a lattice of a few ten sites as the one considered in Fig.\ref{topoevol}, and a Floquet sequence of $N=4$ steps per period as shown in Fig.\ref{hop}, the time-evolution of a few tens of periods that is needed for a complete round-trip around the system is realized via the sequential application of about a hundred two-qubit gate operations in total, a number that is well within reach of currently available devices \cite{google}.

\begin{figure*}[htbp]
    \includegraphics[height = 0.26\linewidth]{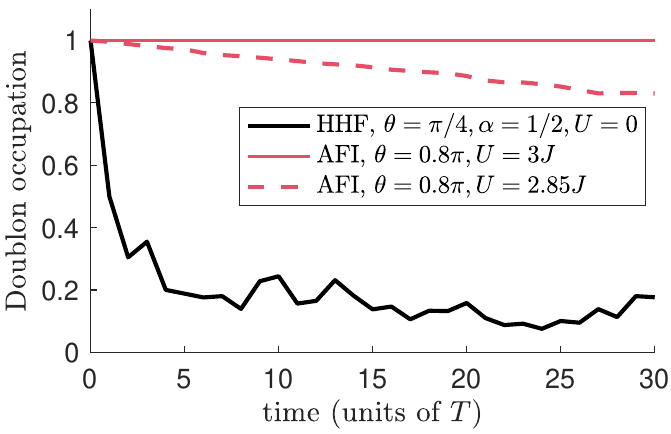}
    \includegraphics[height = 0.26\linewidth]{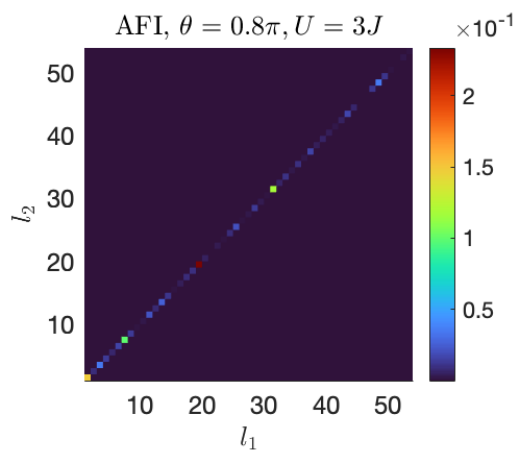}
    \includegraphics[height = 0.26\linewidth]{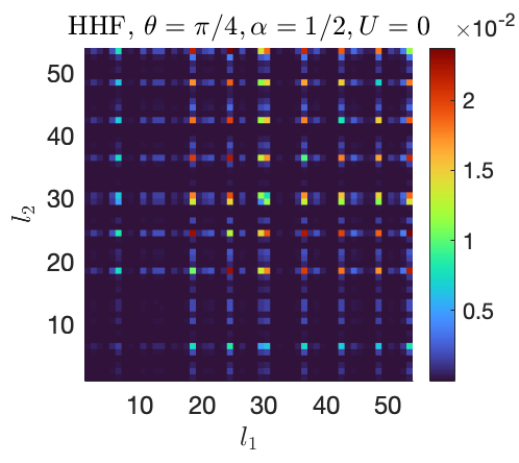}
    \caption{Left panel: time evolution of the overlap
    of the wavefunction with the doublon subspace for the AFI model with two different choices of parameters, exactly satisfying the stability condition \eqref{uj} (solid red) and slightly departing from it by $5\%$ (red dashed). For comparison, the black line shows the same quantity for a non-interacting HHF model with single-particle topology only.  
    Middle and right panels: squared modulus $|\avg{\psi|l_1l_2}|^2$ of the wavefunction, at time $t= 24 T$, in the $(l_1,l_2)$ basis labelled by the position of the two particles. Here, the doublon states correspond to the main diagonal. The middle (right) panels are for the interacting AFI (non-interacting HHF) models. Same system geometry of Fig.\ref{topoevol}.  }
    \label{ent}
\end{figure*}

\subsection{Entanglement features}

The difference between the topological two-particle dynamics and the one of non-interacting single particle edge-states is highlighted by comparing the evolution in Fig.\ref{topoevol} with the one of a non-interacting HHF model at flux $\alpha = 1/2$ with a hopping angle $\theta = \pi/4$ featuring a non-trivial band topology at the single-particle level (right panel of Fig. \ref{spectra}). 
While the spatial evolution of the double-occupancies $A_l$ is qualitatively identical in the two cases (not shown), crucial differences are visible in the left panel of Fig.\ref{ent}. For the topological doublons, the condition \eqref{uj} guarantees that the overlap $\mathcal{O}_{d}=\sum_l A_l$ with the subspace of doublon states remains perfect during the evolution (solid red line) and only suffers minor losses at the end of a round trip even when this condition is approximately satisfied (dashed red line). 
On the other hand, in the single-particle topological case, $\mathcal{O}_{d}$ has already significantly dropped after a few Trotter steps (black line): since there is nothing that can keep the two particles together, their relative wavefunction is free to spread under the effect of the edge state dispersion. 
 
Additional microscopic details of this physics are visible in the middle and right panels of Fig.\ref{ent}. Here, the components of the two-body wavefunction at time $t=24T$ on the basis of two-particle states $\ket{l_1l_2}$ are plotted for the same two cases. 

For non-interacting particles, the wavefunction of the system is the tensor product of the two single-particle states: This is illustrated by the checkerboard pattern visible in the right panel.
In the case of topological doublons, instead, the wavefunction consists of a superposition of doublon states at different positions. In the middle panel, this corresponds to the wavefunction being localized along the $l_1=l_2$ diagonal.

Physically, this behaviour can be understood as a strong entanglement between the two particles: a measurement of the position of the first particle provides full information on the other particle's position. When written in the computational basis of the quantum computer device, the state with a delocalized doublon along the edge can be seen as a generalized W state for a system of $N_\text{sites}$ qutrits. As such, it can be straightforwardly revealed in a QC platform. 

\section{Towards a many-body dynamics}
\label{manybody}
In order to assess the promise of our scheme as a building block for a full-fledged many-doublon physics, we need to check the stability of a pair of doublons located on neighboring sites. As long as the two sites are not coupled, the two doublons do not affect each other and maintain their stability. During the time intervals where the two sites are coupled, undesired processes where the doublons exchange particles and decay into a three body complex plus a free particle or merge into a single four-body complex may occur. 

Assuming for simplicity a two-body form of interactions as included in \eqref{ham}, the interaction energy quickly grows with the number of excitations present. This suggests that the three- and four-body complexes are strongly detuned and their occupation remains small in spite of the interval $\tau$ being relatively short and allowing for sizable departures from energy conservation. This guess is quantitatively validated in the right panel of Fig.\ref{thetaprimo} where we show the probability $P_\text{dec}$ for a pair of neighboring doublons to be decayed at the end of the hopping step for parameters exactly satisfying \eqref{uj} for different $k$'s. In a sizable window of parameter space including our operating point, this probability remains below $10\%$. As it is shown in the Appendix, a further reduced $P_{\rm dec}$ can be obtained by including additional interaction terms in the model. This confirms the potential of our system for studies of many-doublon physics.

\section{Conclusions}
\label{conclusion}
This work analyses the effect of a time-periodic driving on the topological properties of two-dimensional bosonic lattice models with on-site interactions. On the single-particle level, we have shown that a Floquet modulation of the hopping strength can induce a non-trivial topology in systems that would be trivial in a static configuration. Moreover, we have proven the existence of topological bound states of two particles, known as doublons, and formulated conditions for their stability under collision and decay: our framework thus appears promising in view of scaling up towards many-body states. 

The modulation scheme we considered well describes the discreteness of operations typical of digital quantum computers and the models we studied are realizable on already existing devices: our results suggest therefore the potential of these platforms for studies of novel phases of synthetic photonic matter in a Floquet context. 
From the different point of view of quantum information, topological doublons can be interpreted as generalized W states for a system of qutrits: as such, they could be employed to study many-body quantum correlations or to benchmark the performance of a specific quantum computer device in supporting entanglement.

\section{Acknowledgements }
We are grateful to Z. Cian, M. Hafezi, E. Kapit, Z. Jiang, E. Jones, C. Neill, P. Roushan for continuous stimulating exchanges. We acknowledge financial support from the Provincia Autonoma di Trento, the Q@TN initiative, the European Union FET-Open grant ``MIR-BOSE'' (n.737017), from the H2020-FETFLAG-2018-2020 project ``PhoQuS'' (n.820392) and from Google via the quantum NISQ award.

\section*{Appendix: Doublon decay}
In order to assess the robustness of doublons, we compute here the probability that a pair of doublons occupying neighbouring sites merge or exchange particles during a hopping step connecting the two sites. Let us focus on two sites and let us call $a,b$ ($a^\dagger, b^\dagger$) the annihilation (creation) operators for single particles on those sites. Including three- and four-body interactions of strength respectively $U'$ and $U''$, the Hamiltonian reads:
\begin{equation}
\begin{split}
    \mathcal H &= - J \Big( a^\dagger b +  b^\dagger  a \Big) + \frac{U}{2} \sum_{x=a,b}(x^\dagger)^2 (x)^2  +\\
    &\qquad + \sum_{x=a,b} \left[  \frac{U'}{3!} (x^\dagger)^3(x)^3 + \frac{U''}{4!}  (x^\dagger)^4 (x)^4 \right] 
\end{split}
\end{equation}
In our case of four particles distributed among two sites, we can use the following basis of five orthonormal states:
\begin{align*}
    \ket{D} &= \frac{1}{2} a^\dagger a^\dagger b^\dagger b^\dagger\,|\textrm{vac}\rangle \equiv \alpha^\dagger \beta^\dagger\,|\textrm{vac}\rangle\\
    \ket{T_a} &= \frac{1}{\sqrt{3!}} a^\dagger a^\dagger a^\dagger b^\dagger\,|\textrm{vac}\rangle \qquad
    \ket{T_b} = \frac{1}{\sqrt{3!}} a^\dagger b^\dagger b^\dagger b^\dagger\,|\textrm{vac}\rangle\\
    \ket{Q_a} &= \frac{1}{\sqrt{4!}} a^\dagger a^\dagger a^\dagger a^\dagger\,|\textrm{vac}\rangle\qquad
    \ket{Q_b} = \frac{1}{\sqrt{4!}} b^\dagger b^\dagger b^\dagger b^\dagger\,|\textrm{vac}\rangle
\end{align*}
The first one is a pair of doublons sitting at sites $a$ and $b$; the second (third) is a triplet on site $a$ ($b$) plus a single particle in site $b$ ($a$); the fourth (fifth) is a four-body state on site $a$ ($b$). 
Within this basis, the matrix representation of the Hamiltonian is:
\begin{align*}
H = \begin{pmatrix}
2U & -\sqrt{6}J & -\sqrt{6}J & 0 & 0 \\
-\sqrt{6}J & H_T & 0 & -2J & 0 \\
-\sqrt{6}J & 0 & H_T & 0 & -2J \\
0 & -2J & 0 & H_{Q} & 0 \\
0 & 0 & -2J & 0 & H_{Q}
\end{pmatrix}
\end{align*}
where $H_T = U'+3U$ and $H_{Q} = U''+4U'+6U$ are the interaction energies of a three- and four-body state respectively. 
If $M= \exp(-iH\tau)$ the time evolution operator after a trotter step, the probability that the doublons do not remain intact is:
\begin{equation}
    P_\text{dec} = 1-|M_{11}|^2
\end{equation}
The left panel of Fig. \ref{thetaprimo} in the main text shows the trend of $P_\textrm{dec}$ as a function of the effective hopping angle $\theta'$ when $U' = U''=0$ and $U$ is a solution of the decoupling condition Eq.(\ref{uj}) for $k=1,2,3,4$: there exists a sizeble window of values of $\theta'$, close to $\pi/2$, for which the two-body evolution is both topological and robust under doublon decay, because $P_\textrm{dec} \ll 1$. In particular, at our working point (the red dot in Fig. 3), we find $P_\textrm{dec} \simeq 8\%$ while an optimal value around $P_\textrm{dec} \sim 0.4\%$ can be obtained with a better tuned $\theta'\simeq 0.42\pi$. 

\begin{figure}[b]
    \centering
    \includegraphics[height = 0.75\linewidth]{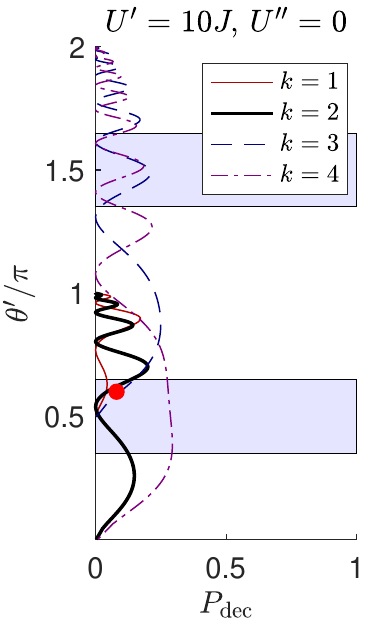}\quad 
    \includegraphics[height= 0.75\linewidth]{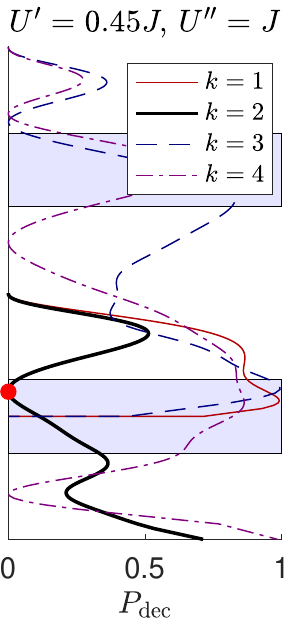}
    \caption{Doublon decay probabilities as a function of the effective hopping angle with $U$ satisfying the the decoupling condition Eq.(\ref{uj}) for $k = 1,2,3,4$. Blue shaded areas are topological for the doublon dynamics. The red dot indicates the working point of Fig.\ref{topoevol} with $\theta' = 0.6\pi$ and $U=3J$. Left panel:  strongly interacting regime, $U' = 10J$, $U''=0$. Right panel: fine tuning procedure with $U' = 0.45J$ and $U''=J$. }
    \label{fig:prob}
\end{figure}

If needed, the stability of doublons can be further improved by considering three- and four-body interactions. The simplest strategy consists in setting $U' \gg J$: we expect such condition to bring triplets and quadruplets out of resonance, leading to a decay probability $P_\textrm{dec}$ closer to 0 for all hopping angles and for any value of $U''$. Indeed, once three-body interactions are sufficiently strong, both $H_T$ and $H_Q$ are already much larger than $J$, without the need of including four-body terms in the Hamiltonian. 
The left panel of Fig. 6 shows the decay probabilities obtained with $U'=10J$ and $U''=0$: the curves remain below $30\%$ for all hopping angles. At our working point, identified again by the red dot, we obtain $P_\textrm{dec} \simeq 4\%$.  Better results are found when further increasing $U'/J$. We verified that adding four-body interactions does not modify significantly the decay probabilities. 

Along the same lines of what we did in the main text for single doublon hopping, we can avoid the strongly interacting regime and proceed to a fine-tuning of $U',U''$ in order to minimize $P_\textrm{dec}$ for all hopping angles. A rigorous procedure to find the optimal values of $U'(\theta),U''(\theta)$ would require the exact analytical computation of the matrix $M$ and the minimization of the coefficients $M_{1l}$, $l\ne1$. Here we restrict ourselves to showing that, once the working point is fixed, one can properly tune $U'$ and $U''$ in order to minimize $P_\textrm{dec}$. This procedure is illustrated in the right panel of Fig. 6 where, for instance, we obtained a decay probability as low as $P_\textrm{dec}(\theta' = 0.6\pi) = 0.02\%$ by tuning $U' = 0.45J$ and $U''=J$ %and optimizing over the value of $\theta'$
.

\nocite{*}
\bibliography{apssamp}

\end{document}